\pdfoutput=1
\documentclass[twocolumn,
floatfix,
preprintnumbers,citeautoscript,prl,superscriptaddress,amsmath,amssymb]{revtex4}

\usepackage{amsmath}
\usepackage{amssymb}
\usepackage[utf8]{inputenc}
\usepackage{graphicx}
\usepackage{hyperref}
\usepackage{color}
\usepackage{silence}

\def\be#1\ee{\begin{align}\begin{split}#1\end{split}\end{align}}

\newcommand{\Mp}{M_{\rm Pl}}

\begin{document}

\title{Phenomenological Consequences of the Refined Swampland Conjecture}

\author{Hajime Fukuda}
\affiliation{Kavli Institute for the Physics and Mathematics of the Universe (WPI), University of Tokyo Institutes for Advanced Study, University of Tokyo, Chiba 277-8583, Japan}

\author{Ryo Saito}
\affiliation{Graduate School of Sciences and Technology for Innovation, Yamaguchi University, Yamaguchi 753-8512, Japan}

\author{Satoshi Shirai}
\affiliation{Kavli Institute for the Physics and Mathematics of the Universe (WPI), University of Tokyo Institutes for Advanced Study, University of Tokyo, Chiba 277-8583, Japan}

\author{Masahito Yamazaki}
\affiliation{Kavli Institute for the Physics and Mathematics of the Universe (WPI), University of Tokyo Institutes for Advanced Study,  University of Tokyo, Chiba 277-8583, Japan}

\date{\today}

\preprint{IPMU-18-0165}

\begin{abstract}
We discuss phenomenological consequences of the recently-introduced refinements of the de Sitter swampland conjecture. The conjecture constrains the first and the second derivatives of the scalar potential in terms of two $O(1)$ constants $c$ and $c'$, leading to interesting constraints on particle phenomenology, especially inflationary model building. Our work can also be regarded as bottom-up constraints on the values of $c$ and $c'$.
\end{abstract}

\maketitle

\bigskip\noindent
{\bf Introduction}

It has been a fascinating question if theories of quantum gravity give rise to 
any novel low-energy constraints beyond those discussed in the framework of the low-energy effective quantum field theory.
To address this question, several ``swampland conjectures'' have been proposed (see e.g. \cite{Brennan:2017rbf} for recent review).
The claim is that these conjectures should be satisfied if the low-energy effective field theory in question has a consistent UV completion with gravity included.

One of the most recent among such swampland conjectures
is the striking conjecture (the so-called de Sitter conjecture) 
by Obied et.\ al.\ \cite{Obied:2018sgi} (see also
\cite{Dine:1985he,Maldacena:2000mw,McOrist:2012yc,Sethi:2017phn,Danielsson:2018ztv} for related discussion). This conjecture states that the scalar potential $V$ in a low-energy effective theory admitting a consistent UV completion with gravity should satisfy the constraint
\be
M_{\rm Pl} \, |\nabla V|>c\, V  .
\label{conjecture_orig}
\ee
Here $c$ is an $O(1)$ constant independent of the choice of the theory (as long as we are in four dimensions), 
and $M_{\rm Pl}=2.4\times 10^{18} \textrm{ GeV}$ is the reduced Planck mass. Subsequently there have been many papers discussing this conjecture \cite{Agrawal:2018own,Dvali:2018fqu,Andriot:2018wzk,Achucarro:2018vey,Garg:2018reu,Lehners:2018vgi,Kehagias:2018uem,Dias:2018ngv,Denef:2018etk,Colgain:2018wgk,Roupec:2018mbn,Andriot:2018ept,Matsui:2018bsy,Ben-Dayan:2018mhe,Damian:2018tlf,Conlon:2018eyr,Kinney:2018nny,Dasgupta:2018rtp,Cicoli:2018kdo, Kachru:2018aqn, Akrami:2018ylq,Murayama:2018lie, Marsh:2018kub, Brahma:2018hrd, Choi:2018rze, Das:2018hqy, Danielsson:2018qpa,Wang:2018duq, Han:2018yrk, Visinelli:2018utg, Moritz:2018ani, Bena:2018fqc,Brandenberger:2018xnf,Brandenberger:2018wbg, Quintin:2018loc, Heisenberg:2018rdu,Gu:2018akj,Heisenberg:2018yae,Brandenberger:2018fdd,Ashoorioon:2018sqb,Odintsov:2018zai,Motaharfar:2018zyb,Kawasaki:2018daf,Hamaguchi:2018vtv,Lin:2018kjm,Ellis:2018xdr,Dimopoulos:2018upl,Das:2018rpg}.

After the initial proposal, several authors
examined the bottom-up consequences of the conjecture, in the context of the Higgs field \cite{Denef:2018etk,Murayama:2018lie,Hamaguchi:2018vtv,Choi:2018rze} and the QCD axion \cite{Murayama:2018lie} (see also \cite{Conlon:2018eyr}).
While these constraints do not necessarily exclude the de Sitter conjecture \eqref{conjecture_orig}, some exotic scenarios seem to be 
inevitable, and one might be tempted to conclude that such scenarios are unlikely.

In view of these results, one natural direction is to weaken/refine the conjecture.
Some proposals along these lines have been made in \cite{Dvali:2018fqu,Garg:2018reu,Andriot:2018wzk,Murayama:2018lie}.
Very recently, in particular, Ooguri et.\ al.\ \cite{Ooguri:2018wrx} proposed a refinement which is closely related with the proposal in \cite{Garg:2018reu},
which states
\be
\Mp\, |\nabla V|>c \, V   \textrm{ or } \,
\Mp^2\, \textrm{min}(\nabla\nabla  V) \leq - c' V.
\label{conjecture}
\ee
Here $c$ and $c'$ are $O(1)$ constants,
and $\textrm{min}(\nabla\nabla  V)$ is the minimal eigenvalue of the Hessian $\nabla_i \nabla_j  V$
in an orthonormal frame. We call this the refined de Sitter conjecture.
This conjecture is obviously weaker than the original conjecture \eqref{conjecture_orig}, but is stronger than the conjecture of \cite[v2]{Murayama:2018lie}, which corresponds to the $c'=0$ case of \eqref{conjecture}:
\be
\Mp\, |\nabla V|>c \, V  \, \textrm{ when } \,
\nabla \nabla  V > 0 ,
\label{conjecture_weaker}
\ee
namely the conjecture applies only when the Hessian is positive definite (i.e.\ $\textrm{min}(\nabla\nabla  V)>0$).

While the authors of \cite{Ooguri:2018wrx} point out the connection of the refined de Sitter conjecture to the distance conjecture \cite{Ooguri:2006in}, it is fair to say that 
the conjecture is still speculative, and it is obviously an important question to find out whether or not there are counterexamples to the conjectures \eqref{conjecture} and \eqref{conjecture_weaker} inside the framework of String/M-theory.

Instead of addressing these questions, in this paper
we set out to discuss some phenomenological consequences of the 
conjectures \eqref{conjecture}, \eqref{conjecture_weaker}.
This is an important point to explore---we can of course continue to weaken the conjecture further so that there are less constraints and less counterexamples, but only at the cost of having less low-energy constraints and less powerful consequences. If we have any hope of deriving reasonably strong constraints on low-energy effective theory, then we clearly need to find a delicate balance.

\bigskip\noindent
{\bf Quintessence}

The stable de Sitter vacua, namely the point where we have
\be 
V>0, \quad \nabla V=0, \quad \nabla \nabla V>0,
\ee
is clearly excluded by the refined de Sitter conjecture \eqref{conjecture}.
This means that the origin of the dark energy should not be the cosmological constant, but rather be the quintessence \cite{Ratra:1987rm,Wetterich:1987fm,Zlatev:1998tr} as in the case of the original conjecture \eqref{conjecture_orig} \cite{Obied:2018sgi,Agrawal:2018own}. For example, the quintessence potential of the form 
\be 
V_Q(Q)= \Lambda_Q^4\, e^{-c_Q \frac{Q}{\Mp}} 
\ee
satisfies the conjecture, as long as $c_Q>c$.

\bigskip\noindent
{\bf Higgs, QCD Axion and All That}

The refined version of the de Sitter conjecture removes bottom-up constraints for the Higgs and the QCD axion pointed out in the literature \cite{Denef:2018etk,Conlon:2018eyr,Murayama:2018lie,Hamaguchi:2018vtv,Choi:2018rze}.  While this is technically an easy consequence, it is worth emphasizing this point,
since one of the important motivations for the refinements of the de Sitter conjectures \eqref{conjecture}, \eqref{conjecture_weaker}
is to evade the bottom-up constraints.

The potential for the Higgs field $H_{\rm SM}$
\be 
V_{H_{\rm SM}}(H_{\rm SM})=\lambda (|H_{\rm SM}|^2-v^2)^2 
\ee
has a local maximum at $H_{\rm SM}=0$, which violates the first condition in \eqref{conjecture} for the first derivative. However, the second derivative
is non-zero around this point,
where we have
\be 
\Mp^2 \frac{V''}{V} \sim  -O\left( 
   \frac{\Mp^2}{(100 ~{\rm GeV})^2}
   \right) \ll -c' .
\ee
The similar argument applies to the QCD axion, which has a cosine potential,
or more generally many scenarios for spontaneous symmetry breaking.
In fact, as these examples show, the condition \eqref{conjecture}
is satisfied for a generic potential for energy scale much smaller than the Planck scale, except when there is an (nearly) stable de Sitter vacua.

\bigskip\noindent
{\bf Inflation}
 
The new conjecture \eqref{conjecture_2} has more non-trivial implications on inflation, for which the potential is often fine-tuned.

For simplicity let us consider a single-field inflation with the canonical kinetic term. Let us denote the inflaton by $\phi$. It is customary to 
define two slow-roll parameters $\epsilon_V$ and $\eta_V$ by
\be
\epsilon_V=\frac{\Mp^2}{2} \left(\frac{V'}{V}\right)^2 , \quad
\eta_V=\Mp^2 \left(\frac{V''}{V}\right),
\ee
and then the refined dS conjecture \eqref{conjecture} reads
\be
[\epsilon]: \,\epsilon_V \geq \frac{c^2}{2} \quad \textrm{ or } \quad
[\eta]: \,\eta_V \leq -c'.
\label{conjecture_2}
\ee
In the slow-roll regime $\epsilon_V \ll 1\,, |\eta_V| \ll 1$, the slow-roll parameters are directly related with the scalar
spectral index $n_s$ and the tensor-to-scalar ratio $r$
by the relation,
\be
n_s=1-6\epsilon_V+2 \eta_V,\quad
r=16 \epsilon_V.
\ee

The parameters $c$ and $c'$ are bounded by the conjecture for consistency with a  canonical single-field inflation.
If we try to satisfy the first condition $[\epsilon]$ in \eqref{conjecture_2},
then we have $r\geq 8 c^2$. By combining this with the current observational bound $r<0.064$ \cite{Akrami:2018odb} we have $c<0.09$, 
which is in some tension with the conjecture, as already pointed out in \cite{Agrawal:2018own, Achucarro:2018vey, Garg:2018reu, Kehagias:2018uem, Dias:2018ngv,Kinney:2018nny}. 
The inflation also relates the condition $[\epsilon]$ to another conjecture \cite{Achucarro:2018vey,Dias:2018ngv}, the distance conjecture \cite{Ooguri:2006in} $\Delta \phi/\Mp < \Delta$ with a constant $\Delta = O(1)$.
The restriction is relaxed in the new conjecture \eqref{conjecture_2}:
\be
	\frac{\Delta \phi}{\Mp} = \int \sqrt{2\epsilon_V} {\rm d}N_e > c N^{\rm (convex)}_e ,
\ee
where $N_e^{\rm (convex)}$ is the e-fold elapsed in the convex region of the potential, 
instead of that measured from the end of inflation.

In the refined version of the conjecture \eqref{conjecture}
we have another option, namely to satisfy the condition $[\eta]$ during inflation.
This in particular implies that the inflaton potential should be concave ($\eta_V<0$), and this is also favored by recent observations. 
The condition $\eta_V \leq -c'$ immediately leads to 
the bound on the size of the tensor-to-scalar ratio,
\begin{align}
r\leq \frac{8}{3}(1-2c'-n_s).
\label{r_bound}
\end{align}
Taking into account the scalar modes as well as the tensor modes, 
the observational data lead to $\epsilon_V < 0.005\,, \eta_V \simeq -0.01$ without assuming the slow-roll approximation \cite{Akrami:2018odb}.
We then have $c' < 0.01$.

A small value of $c'$ is also indicated by another argument, 
namely to get a sufficiently long period of inflation, $N_e = 50 \-- 60$.
The condition $[\eta]$ is not a direct obstacle for inflation, 
but $\epsilon_V$ should be tuned to a tiny value
because the slope of the potential rapidly increases as the inflaton rolls down the potential. 
However, in this case, 
the energy scale of the inflation becomes very low 
to give the observed value of the scalar power spectrum amplitude $A_s \propto V/\epsilon_V$,
which conflicts with the lower bound on the reheating temperature from Big Bang Nucleosynthesis \cite{deSalas:2015glj} when $c'$ is $O(1)$.

It is noteworthy that, in contrast to the bound on $c$, the observational data on the scalar modes is used to get the bound on $c'$ \footnote{See however \cite{Hetz:2016ics,Covi:2008cn}.}.
Hence, the latter bound depends on the generation mechanism of the curvature perturbations. For example, in the curvaton scenario \cite{Enqvist:2001zp,Lyth:2001nq,Moroi:2001ct}, the scalar spectral index is given by,
\be
	n_s = 1 + 2\frac{\dot{H}}{H^2} + 2\eta_{\sigma\sigma} \,,
    \label{eq:curvaton}
\ee
for the curvaton $\sigma$ with $\eta_{\sigma\sigma} \equiv V_{\sigma\sigma}/3H^2$, where the quantities are evaluated at the horizon exit. 
When there is no coupling with the inflaton, the conjecture (\ref{conjecture}) reads $|\eta_{\sigma\sigma}| > c' \Omega_\sigma$ 
for the energy fraction of the curvaton $\Omega_\sigma$ at the horizon exit. 
Thus, $\eta_{\sigma\sigma}$ can be small when the curvaton is sufficiently subdominat during inflation.
The tiny value of $\epsilon_V$ for a sufficiently long period of inflation leads to $n_s \simeq 1 + 2\eta_{\sigma\sigma}$. 
Consistency with the observed value of $n_s$ requires $\eta_{\sigma\sigma}\sim  -O(0.01)$.
This indicates that the curvaton should have a potential with a hilltop region. Such curvaton models can be consistent with observations, however some specific models such as a pseudo-Nambu-Goldstone curvaton generates too much non-Gaussiniaty \cite{Kawasaki:2011pd}.

\bigskip\noindent
{\bf Global Constraints on Inflaton Potential}

We have so far discussed the region of the inflaton potential for the accelerated expansion.
The condition \eqref{conjecture}, however, applies to any 
point in the configuration space (as long as the low-energy effective theory is valid), and hence we need to make sure that there are no extra 
constraints in other regions of the configuration space for the inflaton.

In inflationary models one often assumes that there is a region of the inflaton
potential where the inflaton oscillates around the bottom of the potential and the reheating happens. The potential in this region can for example be taken to be quadratic
(where we have chosen the origin $\phi=0$ to be the minimum of the inflaton potential)
\be
V(\phi) \sim \frac{m^2 \phi^2}{2}.
\ee
Near the origin ($|\phi| < \Mp$) we have $\Mp V'/V= M_{\rm Pl}/\phi$,
and the first derivative condition $[\epsilon]$ is automatically satisfied
as long as $\phi\ll \Mp/c$.

The situation is more non-trivial when the value of the inflaton
becomes large, $\phi\sim \Mp/c$. 
Let us choose the value of the inflaton to be $\phi=\phi_*$
such that we have $V''/V =-c'$. Then 
we need to satisfy 
\be
\Mp \frac{V'}{V} \Big|_{\phi=\phi_*} >c.
\ee
This gives extra non-trivial constraints on the 
parameters of the inflaton potential.
Equivalently, once we fix the inflaton potential
we can use this condition as constraints on the parameters $c$ 
and $c'$. In the following we will sometimes
simplify the analysis by taking $c'=0$.
This corresponds to the weaker conjecture \eqref{conjecture_weaker},
and gives the more conservative estimate of the constraints. 

\begin{figure}[htbp]
\centering\includegraphics[scale=0.3]{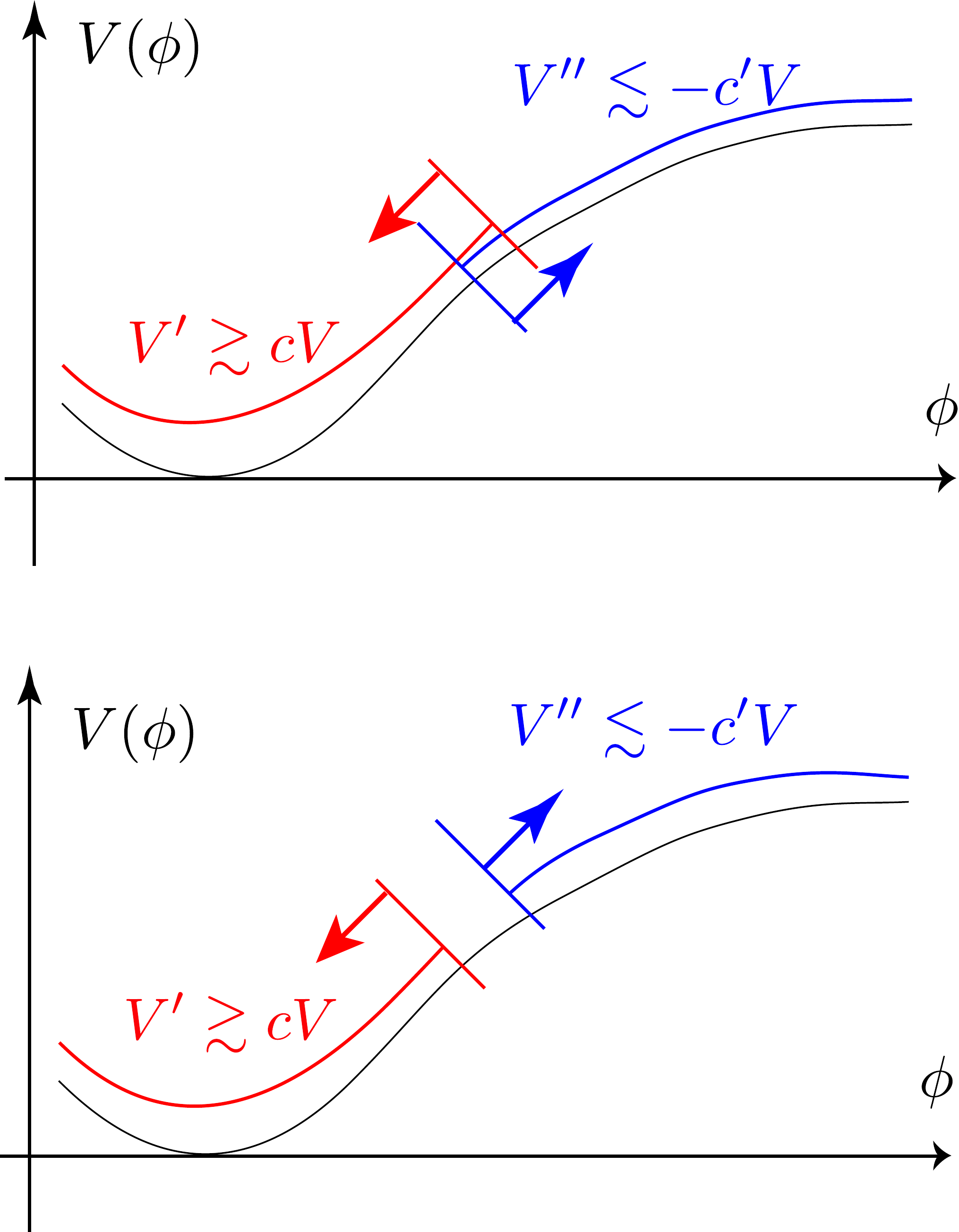}
\caption{For a plateau-type inflaton potential, we find that (1) near the bottom of the potential for reheating the condition $[\epsilon]$ is satisfied (region colored red),
and (2) the condition $[\eta]$ is satisfied in the regions for accelerated expansion (region colored blue). Depending on the choice of the parameters
there could be an intermediate region where neither condition is satisfied, as in the figure below.}
\label{fig.schematic}
\end{figure}

One might notice that in many inflationary models (including pure natural inflation and $\alpha$-attractors discussed below) have plateau, where one might run into 
contradictions with the conjecture \eqref{conjecture}.
Since we have argued below \eqref{r_bound} that $c'<0.01$,
this would happen at large values of the inflaton, where the effective field theory might break down according to the distance conjecture \cite{Ooguri:2006in}.
Our analysis below is conservative in that we will not take this point fully into account.

\bigskip\noindent
{\bf Natural Inflation}

For our further discussion we need to have the inflation scenario where
we know the global form of the inflaton potential.
A good example is provided by the natural inflation \cite{Freese:1990rb,Adams:1992bn}. In this scenario the inflaton potential
is generated by the coupling of the inflaton to the non-Abelian Yang-Mills gauge field,
and the inflaton potential takes the cosine form, as 
generated by the one-instanton:
\be
V(\phi)=V_0 \left(1-\cos\frac{\phi}{f} \right) .
\label{V:natural}
\ee
Here $f$ is the decay constant of the axion and the overall scale $V_0$ is determined by the dynamical scale of the confined Yang-Mills field.

\begin{figure}[htp]
\includegraphics[width=0.45\textwidth]{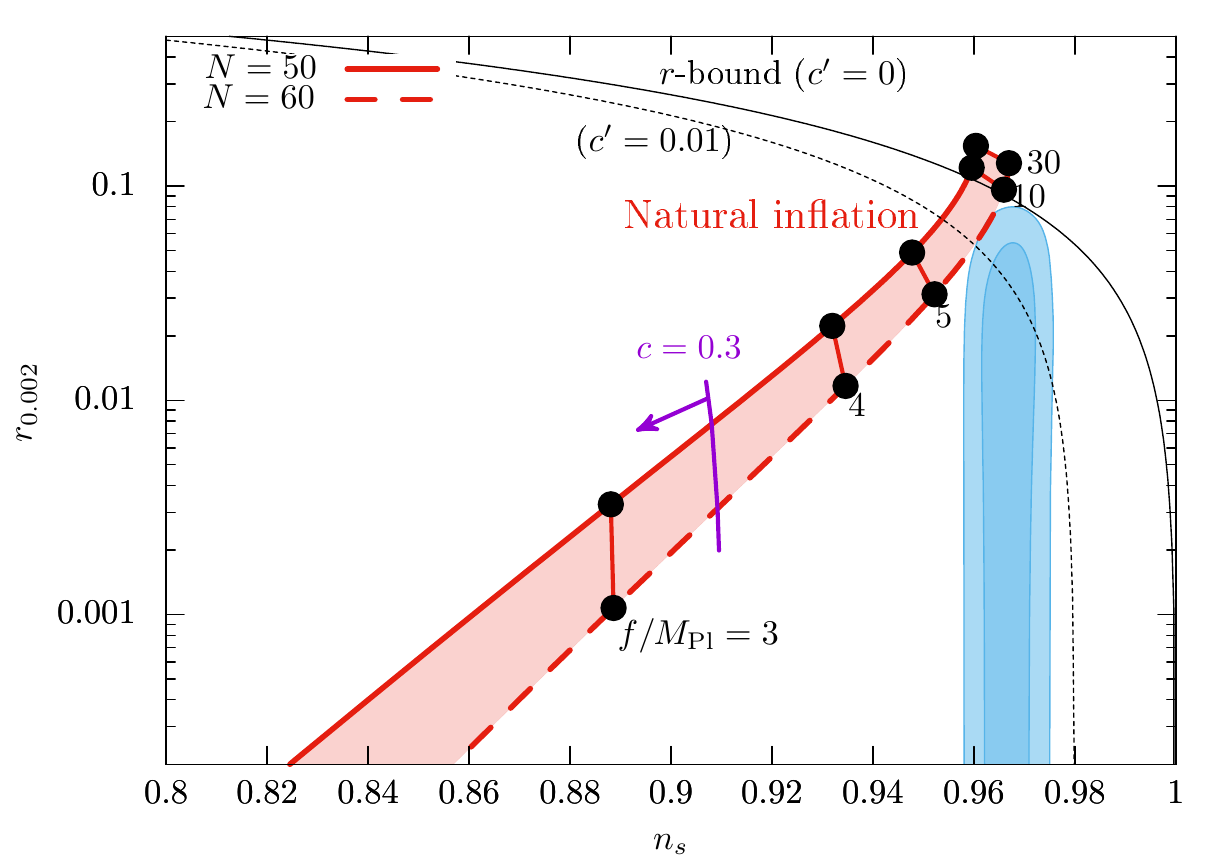}
\caption{The $n_s-r$ plane for the natural inflation \eqref{V:natural}, with e-folding between $50$ and $60$. The constraint \eqref{result:natural} for $c=0.3$
excludes the region below the purple line, making the model inconsistent with current observational constraints \cite{Akrami:2018odb} (blue regions, 68\% and 95\% CL).
The black and solid (dotted) line shows the upper-bound for $r$ \eqref{r_bound} with $c'=0~(0.01)$.
}
\label{fig:natural}
\end{figure}

This potential is $2\pi f$-periodic, has a maximum at $\phi=\pi f$,
and is convex when $|\phi|< \pi f/2$.
In this range the minimal value of the ratio $M_{\rm Pl}V'/V$ should be larger than the $O(1)$ coefficient $c$, so that
\be
M_{\rm Pl}\frac{V'}{V} = \frac{M_{\rm Pl}}{f} \cot\frac{\phi}{2f} \gtrsim \frac{M_{\rm Pl}}{ f} >c,
\ee
namely we obtain an upper bound on the decay constant
\be
f < \frac{M_{\rm Pl}}{c}.
\label{result:natural}
\ee
For $c\gtrsim 1$
this constraint can be stronger than the constraint from the weak gravity conjecture \cite{ArkaniHamed:2006dz}, 
which imposes $f\lesssim O(M_{\rm Pl})$.
Note that we obtain a similar constraint $f< \Mp/\sqrt{c'}$ by applying the second-derivative condition $[\eta]$ near the top of the potential \cite{Ooguri:2018wrx}.
We can also  plot the constraint on the $n_s-r$ plane, see Fig.~\ref{fig:natural} 
for exclusion region for $c=0.3$.

\begin{figure}[htp]
\includegraphics[width=0.45\textwidth]{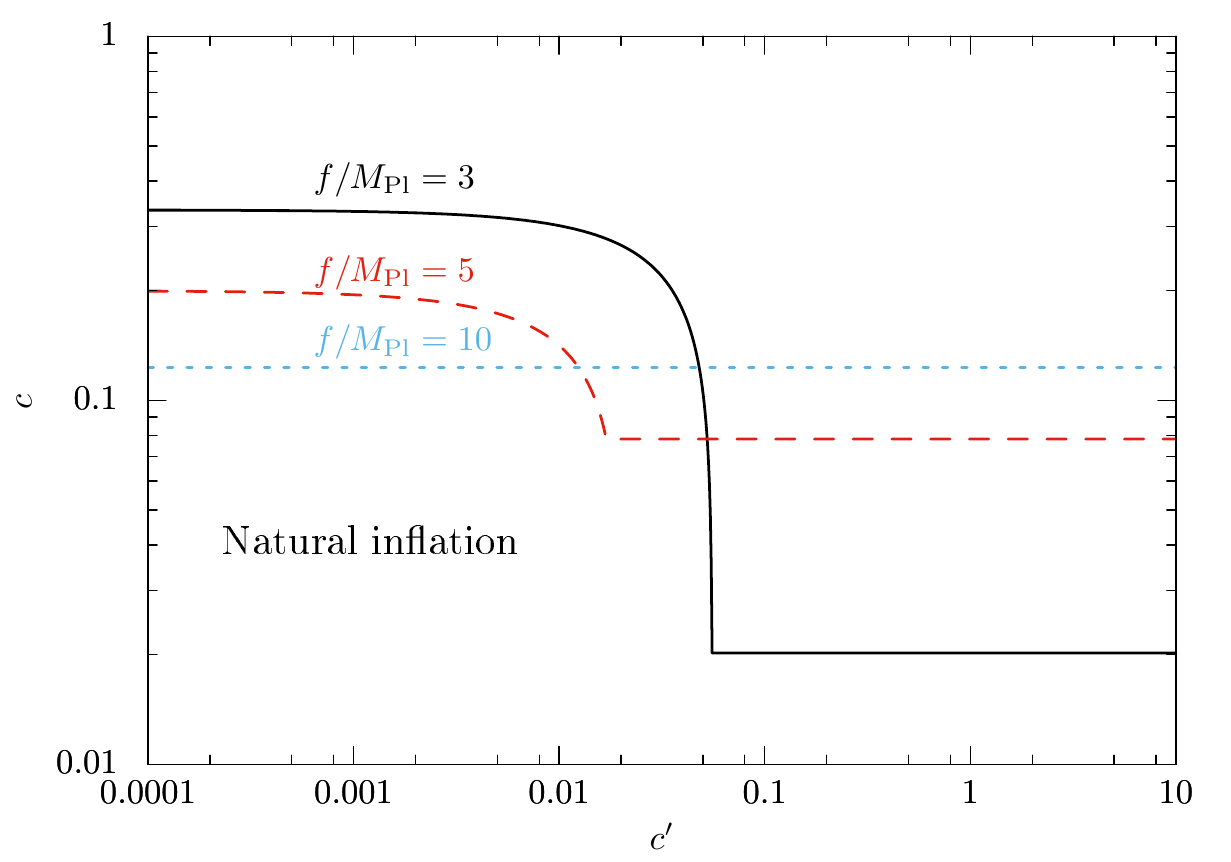}
\caption{The constraint of $c$ and $c'$ for the natural inflation model with e-folding 50.
The regions below the lines are allowed.
}
\label{fig:natural_cc}
\end{figure}

We have chosen $c'=0$ above for simplicity, 
but we can analyze the more general case $c'>0$. 
We impose consistency with the conjectures for the inflaton potential
in the whole region between the initial field value for inflation and the global minimum. In Fig.~\ref{fig:natural_cc}, we show the resulting constraints for $c'$ and $c$ for the natural inflation with e-folding $50$.

\bigskip\noindent
{\bf Pure Natural Inflation}

\begin{figure}[htp]
\includegraphics[width=0.45\textwidth]{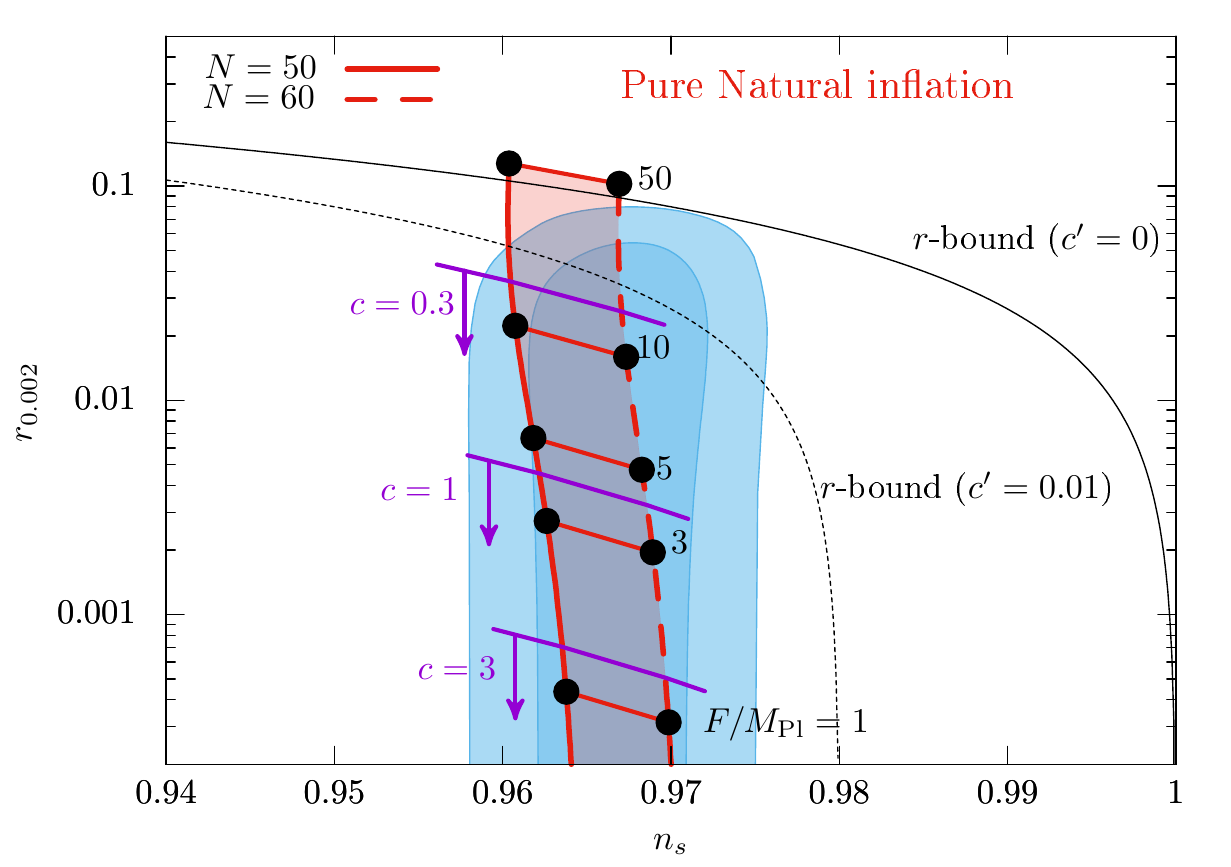}
\caption{
Same as in Fig.~\ref{fig:natural} but for the pure natural inflation \eqref{V:p_natural}, with e-folding between $50$ and $60$. The constraint \eqref{result:p_natural} for $c=0.3,1,3$ excludes the region below the purple line. }
\label{fig:p_natural}
\end{figure}

\begin{figure}[htp]
\includegraphics[width=0.45\textwidth]{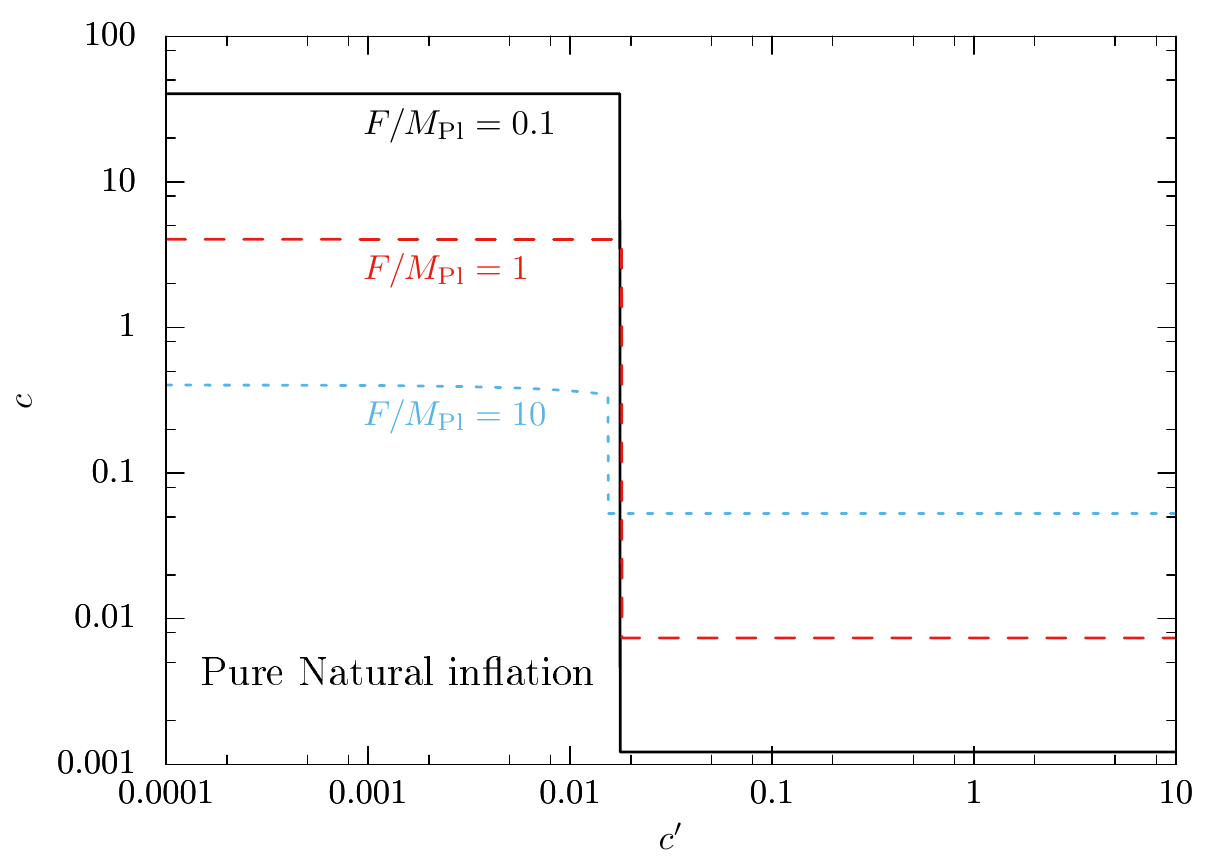}
\caption{Same as in Fig.~\ref{fig:natural_cc} but for the pure natural inflation.
}
\label{fig:p_natural_cc}
\end{figure}

We can use the inflation model
which takes advantage of the multiple-branch structure of the 
potential as a function of the $\theta$-angle \cite{Nomura:2017ehb} (this is the field theory version of the monodromy inflation  \cite{Silverstein:2008sg,McAllister:2008hb}).
The potential reads
\be
V(\phi)=V_0 \left[ 1-
\left(1+\left(\frac{\phi}{F}\right)^2\right)^{-p}
\right],
\label{V:p_natural}
\ee
where $V_0$ is again determined by the dynamical scale of the Yang-Mills
field and $F$ is the effective decay constant.
The power $p$ in the potential \eqref{V:p_natural} 
is a positive integer, which could be 
determined by future improved lattice simulations \cite{Nomura:2017zqj}.
Let us here choose the holographic value $p=3$ \cite{Dubovsky:2011tu}.
We have $V''(\phi)=0$ at $\phi/F=1/\sqrt{7}$.
The ratio $M_{\rm Pl} V'/ V$ is a rapidly decaying function,
whose minimum is attained at $\phi/F=1/\sqrt{7}$.
Requiring that this value is greater than $c$, we obtain
\be
\frac{F}{\Mp} < \frac{1029 \sqrt{7} }{676\,  c } \sim \frac{4.0}{c },
\label{result:p_natural}
\ee
which is shown in Fig.~\ref{fig:p_natural}.
We also show more general constraints of $c$ and $c'$ in Fig.~\ref{fig:p_natural_cc}.

Contrary to the case of the natural inflation,
the pure natural inflation utilizes the multi-branch structure of the potential,
and this raises the question of whether to impose the conjectures \eqref{conjecture} and \eqref{conjecture_weaker} for each metastable branch, or for the branch with the lowest energy.
If we interpret the conjecture in the latter way, then there actually are no constraints of the type discussed here.
See section 8 of \cite{Murayama:2018lie} for related discussion.

\bigskip\noindent
{\bf Starobinsky Model}

Let us next discuss the $R^2$ inflation model by Starobinsky \cite{Starobinsky:1980te}. When we choose the canonical kinetic term for the inflaton, the inflaton potential is given by
\be
V_S(\phi)=V_0 \left(1-e^{-\sqrt{\frac{2}{3}} \frac{\phi}{M_{\rm Pl}}} \right)^{2}.
\label{Starobinsky}
\ee
By repeating the computation as before, we find constraints from the
intermediate regions as in Fig.~\ref{fig.schematic} to be
\be
c < 2 \sqrt{\frac{2}{3}}\sim 1.6.
\ee

\bigskip\noindent
{\bf $\alpha$-attractor}

The Starobinsky model can be thought of as a special case of the 
more general models known as the $\alpha$-attractor
\cite{Kallosh:2013hoa, Galante:2014ifa}.

\begin{figure}[htp]
\includegraphics[width=0.45\textwidth]{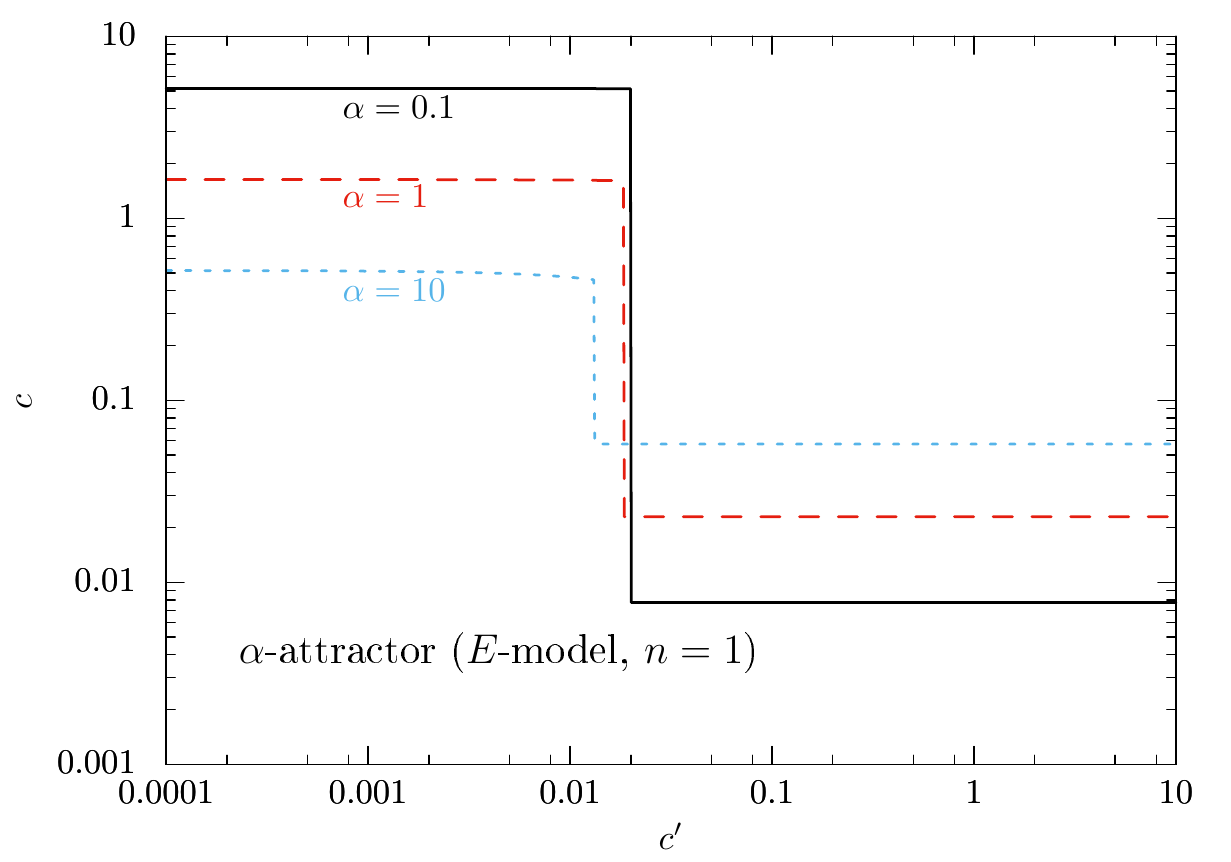}
\caption{Same as in Fig.~\ref{fig:natural_cc} but for the $E$-model $\alpha$-attractor \eqref{V:E} with $n=1$.
}
\label{fig:e_model_cc}
\end{figure}

\begin{figure}[htp]
\includegraphics[width=0.45\textwidth]{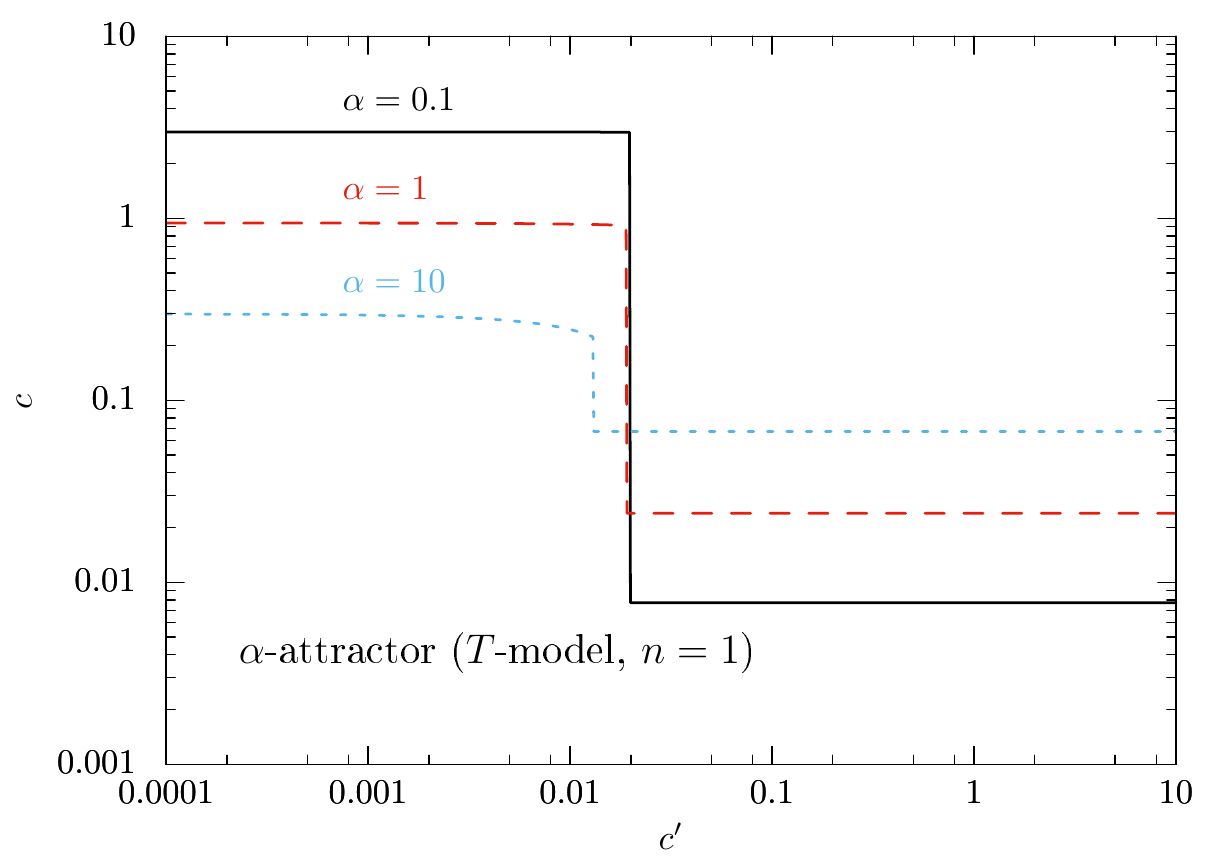}
\caption{Same as in Fig.~\ref{fig:natural_cc} but for the $
T$-model $\alpha$-attractor \eqref{V:T} with $n=1$.
}
\label{fig:t_model_cc}
\end{figure}

The so-called $E$-model of the $\alpha$-attractor is
\be
V_E(\phi)=V_0 \left(1-e^{-\sqrt{\frac{2}{3\alpha}} \frac{\phi}{M_{\rm Pl}}} \right)^{2n},
\label{V:E}
\ee
which includes the Starobinsky model \eqref{Starobinsky}
as a special case $n=1, \alpha=1$.
In this case the constraint on $c$ gives
\be
c < 2 \sqrt{\frac{2}{3 \alpha}} \frac{n}{2n-1} .
\label{result:E}
\ee
We also have the so-called $T$-model
\be
V_T(\phi)=V_0 \tanh^{2n}\left(
\frac{\phi}{\sqrt{6\alpha} M_{\rm Pl}}
\right).
\label{V:T}
\ee
This gives the constraint
\begin{align} 
c<\frac{2\sqrt{2}}{3\sqrt{\alpha}} \quad (n=1), \qquad
c<\frac{4 \sqrt{2}}{3\sqrt{5\alpha}} \quad (n=2).
\label{result:T}
\end{align}
The constraints \eqref{result:E} and \eqref{result:T} are strong
for larger values of $\alpha$, such as $\alpha\sim O(10^2)$,
which are still marginally consistent with the current observations.
Note that the constraints from the field range ($\Delta \phi/\Mp<O(1)$) is also stronger for 
larger values of $\alpha$.
We also show the constraints of $c$ and $c'$ in Figs.~\ref{fig:e_model_cc}, \ref{fig:t_model_cc}.

\begin{figure}[htp]
\includegraphics[width=0.45\textwidth]{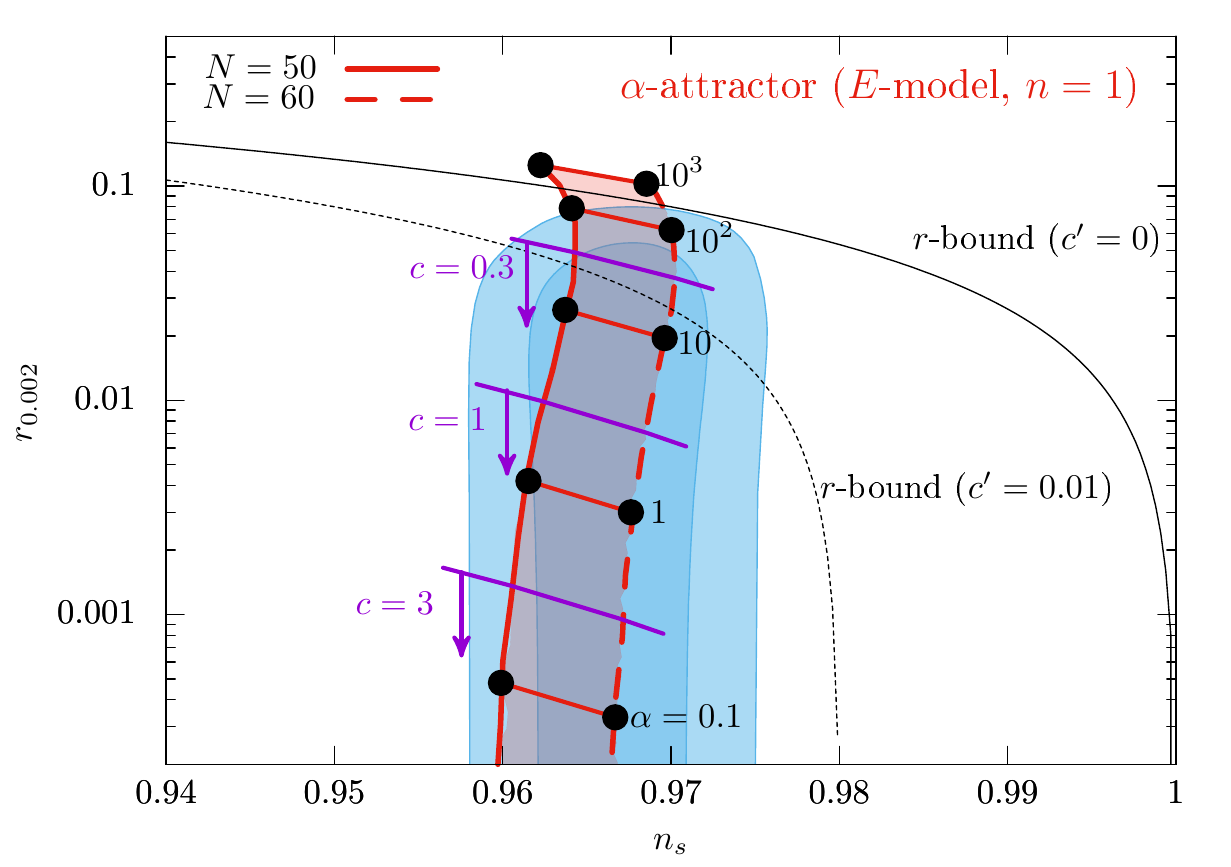}
\caption{Same as in Fig.~\ref{fig:natural} but for the $E$-model $\alpha$-attractor \eqref{V:E} with $n=1$, with e-folding between $50$ and $60$. This model contains the Starobinsky model
\eqref{Starobinsky} as a special case $\alpha=1$.
The constraint \eqref{result:E} for $c=0.3,1,3$ excludes the region below the purple line. }
\label{fig:e_model}
\end{figure}

\begin{figure}[htp]
\includegraphics[width=0.45\textwidth]{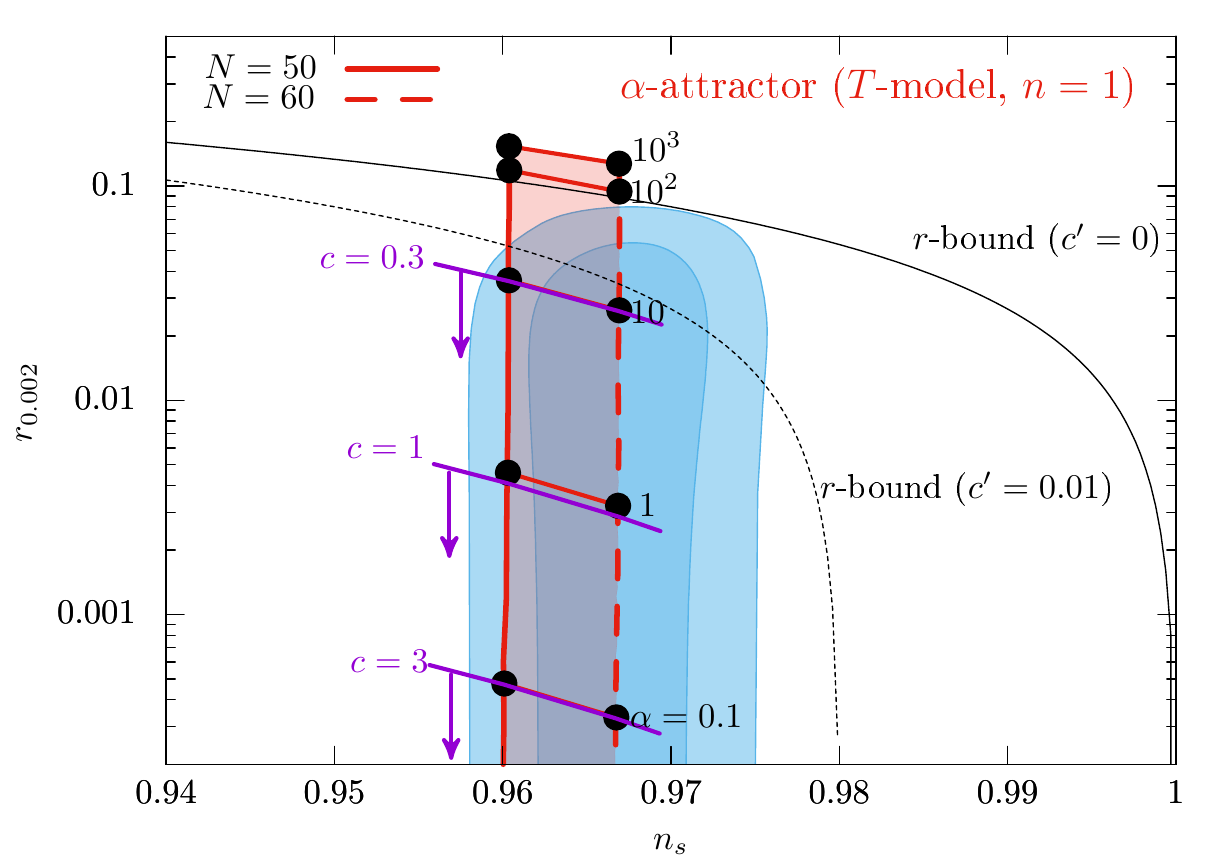}
\caption{Same as in Fig.~\ref{fig:natural} but for the  $T$-model $\alpha$-attractor \eqref{V:T} with $n=1$, with e-folding between $50$ and $60$. The constraint \eqref{result:T} for $c=0.3,1,3$ excludes the region below the purple line.}
\label{fig:t_model}

\end{figure}

\bigskip\noindent
{\bf Acknowledgements}

We would like to thank M.~Ibe, M.~Kawasaki and T.T.~Yanagida for related discussions. This research was supported in part by WPI Research Center Initiative, MEXT, Japan (HF, SS, MY), by JSPS Research Fellowship for Young Scientists (HF), 
and by the JSPS Grant-in-Aid for Scientific Research No.~17K14286 (RS), 17H02878 (SS), 18K13535 (SS), and 17KK0087 (MY).

\bibliographystyle{apsrev4-1}
\bibliography{swampland_2018_10}
\end{document}